# Thermal Conductivity of Oxide Tunnel Barriers in Magnetic Tunnel Junctions Measured by Ultrafast Thermoreflectance and Magneto-optic Kerr Effect Thermometry

Hyejin Jang[1†*], Luca Marnitz[2], Torsten Huebner[2], Johannes Kimling[1], Timo Kuschel[2], and David G. Cahill[1†]

[1] Department of Materials Science and Engineering and Materials Research Laboratory,
   University of Illinois, Urbana, IL 61801, USA

[2] Center for Spinelectronic Materials and Devices, Department of Physics, Bielefeld University,
   Universitätsstrasse 25, 33615 Bielefeld, Germany

[†] Corresponding authors: hjang@berkeley.edu, d-cahill@illinois.edu
[*] Present address: Department of Electrical Engineering and Computer Sciences, University of California, Berkeley, CA 94720, USA

Abstract

Spin-dependent charge transport in magnetic tunnel junctions (MTJs) can be manipulated by a temperature gradient, which can be utilized for spintronic and spin caloritronic applications. Evaluation of the thermally induced phenomena requires knowledge of the temperature differences across the oxide tunnel barrier adjacent to the ferromagnetic (FM) leads. However, it is challenging to accurately measure thermal properties of an oxide tunnel barrier consisting of only a few atomic layers. In this work, we experimentally interrogate the temperature evolutions in Ru/oxide/FM/seed/MgO (oxide=MgO, MgAl$_2$O$_4$; FM=Co, CoFeB; seed=Pt, Ta) structures having perpendicular magnetic anisotropy using ultrafast thermometry. The Ru layer is optically thick and heated by ultrafast laser pulses; the subsequent temperature changes are monitored using thermoreflectance of Ru and magneto-optic Kerr effect (MOKE) of the FM layers. We independently measure the response times of Co and CoFeB magnetism using quadratic MOKE and obtain $\tau_{em}$=0.2 ps for Co and 2 ps for CoFeB. These time scales are much shorter than the time scale of heat transport through the oxide tunnel barrier, which occurs at 10−3000 ps. We determine effective thermal conductivities of MgO and MgAl$_2$O$_4$ tunnel barriers in the range of 0.4−0.6 W m$^{-1}$ K$^{-1}$, comparable to an estimate of the series conductance of the Ru/oxide and oxide/FM interfaces and an order of magnitude smaller than the thermal conductivity of MgO thin films. We find that the electron-phonon thermal conductance near the tunnel barrier is only a factor of 5−12 larger than the thermal conductance of the oxide tunnel barrier. Therefore, the drop in the electronic temperature is approximately 20−30% larger than the drop in the phonon temperature across the tunnel barrier.

## I. INTRODUCTION

Magnetic tunnel junctions (MTJs) are essential components of information technology. They are used as read heads for hard disk drives and one MTJ stores a bit of data in spin-transfer-torque magnetoresistive random access memory (STT-MRAM), which is in its early stage for commercial production [1,2]. The key feature of MTJs that enables these applications is that they show tunnel magnetoresistance (TMR) effect at room temperature, i.e., a large contrast in electrical resistance depending on the magnetic configuration of the MTJ. For MTJs consisting of CoFeB/MgO/CoFeB trilayers, TMR ratios up to 600% have been observed at room temperature [1,3].

Passing a heat current through an MTJ leads to further phenomena in which heat, charge, and spin transport are coupled, thereby increasing the number of potential spintronic applications. For example, the tunnel magneto-Seebeck (TMS) effect [4,5] refers to the change of the Seebeck coefficient of an MTJ depending on its magnetic configuration. Together with additional magnetothermoelectric effects, the TMS can be utilized, e.g., for three-dimensional sensing of temperature gradients in nanostructures. [6]. Application of a heat current to an MTJ also induces a thermal spin-transfer torque, which can assist magnetic switching [7–10]. And Seebeck spin tunneling occurs in an oxide tunnel barrier in contact with a ferromagnetic (FM) metal and a non-magnetic semiconductor, allowing for spin current injection into a semiconductor [11].

The analysis of all these thermally driven phenomena requires knowledge of the temperature differences inside the tunneling devices, and thus knowledge of the thermal transport properties of the tunneling devices. MTJs are usually composed of metallic materials except for an oxide tunnel barrier. Some general aspects of thermal transport properties of metals are as follows.

The Wiedemann-Franz law in Eq. (1) relates electrical conductivity ($\sigma$) to electronic contribution of the thermal conductivity ($\Lambda_e$).

$$\Lambda_e = \sigma L T \qquad (1)$$

$L$ is the Lorenz number and the Sommerfeld theory gives $L=2.44\times10^{-8}$ W $\Omega$ K$^{-2}$. For example, the intrinsic electrical resistivities of Cu and Co are 1.54 µΩ cm and 5.2 µΩ cm, respectively, at 300 K [12], which converts into thermal conductivities of $\Lambda_e \approx$ 480 W m$^{-1}$ K$^{-1}$ for Cu and 140 W m$^{-1}$ K$^{-1}$ for Co. The interface between different materials represents a discontinuity of materials and substantially disrupts heat transport. For metal-metal interfaces, electronic thermal transport dominates. For example, the interface between sputtered Al and Cu films shows a thermal conductance ($G$) of 4 GW m$^{-2}$ K$^{-1}$ [13], corresponding to an effective thermal conductivity of 4 W m$^{-1}$ K$^{-1}$ as an 1-nm-thick layer analogue. Multilayers of thin metals show suppressed thermal conductivities due to i) boundary scattering of electrons as the layer thicknesses become comparable to the mean-free-paths of electrons (1–10 nm) and ii) the increased density of interfaces. For the [Co(1.2 nm)/Cu(1.1 nm)]$_{180}$ multilayer, where 180 is the repetition number, $\Lambda_e \approx$ 5–7 W m$^{-1}$ K$^{-1}$ is derived from perpendicular magnetoresistance measurements [14].

In dielectric materials, phonons dominate heat transport. For MgO, the bulk thermal conductivity of 48 W m$^{-1}$ K$^{-1}$ [15] is reduced to 4 W m$^{-1}$ K$^{-1}$ in nanostructured films having grain sizes of 3-7 nm [16]. The metal-dielectric interface interrupts heat transfer more significantly than the metal-metal interface [17–20]. For example, multilayers of W/Al$_2$O$_3$ nanolaminates show strongly reduced thermal conductivity of 0.6–1.5 W m$^{-1}$ K$^{-1}$ compared to the thermal conductivities of each W and Al$_2$O$_3$ layer. The major contribution to this suppression comes from the W/Al$_2$O$_3$ interface having $G = 0.26$ GW m$^{-2}$ K$^{-1}$ [17]. The upper limit of thermal conductance of metal-dielectric interface is explored in Ref. [20]: for the Al/MgO interface, the clean interface has $G = 0.5$ GW m$^{-2}$ K$^{-1}$ at ambient pressure, which increases to 1 GW m$^{-2}$ K$^{-1}$ under the pressure of 60 GPa. At the metal-dielectric interface, the heat current is mainly carried by phonons. Although the remote coupling between electrons and phonons across a metal-dielectric interface has been suggested as a possible channel for interfacial heat transport, experiments indicate that the role is limited [18,21].

In tunneling devices, the heat current from electronic transport through the tunnel barrier is negligible relative to the heat current carried by phonons. For example, the CoFeB/MgO/CoFeB MTJ in Ref. [22] has a resistance-area product ($RA$) of 3 Ω μm$^2$ in the parallel state. The corresponding thermal conductance of tunneling electrons ($G_e$) according to Eq. (1) is $G_e = LT/RA \approx 2$ MW m$^{-2}$ K$^{-1}$. Although Ref. [23] reported the deviation from the Wiedemann-Franz law in MTJs due to vacancy defects, the change in the Lorenz number is 30% at most at 300 K. Thus, the Wiedemann-Franz law still provides a reasonable estimate for $G_e$. As we show below, $G_e \approx 2$ MW m$^{-2}$ K$^{-1}$ is much smaller than the thermal conductance of phonons ($G_{ph}$) through the tunnel barrier, which is of the order of 100 MW m$^{-2}$ K$^{-1}$.

Therefore, we expect that in MTJs under a temperature gradient, the oxide tunnel barrier of 1-2 nm thickness and its interfaces with FM metals, e.g., CoFeB, possess the smallest effective thermal conductivity conductance among the other components, i.e., the largest temperature difference in MTJs occurs at the oxide tunnel barrier.

The first experimental work reporting the TMS effect [4] adopted a thermal conductivity of the nanostructured MgO thin films from Ref. [16], $\approx 4$ W m$^{-1}$ K$^{-1}$, to assess the TMS performance. However, several theoretical [24] and experimental [25,26] studies suggested that the effective thermal conductivity of an oxide tunnel barrier in MTJs can be much smaller, about an order of magnitude, than the thin film value of Ref. [16], which indicates that the size of the TMS effect determined was overestimated. Zhang *et al.* [24] used a Green function approach to calculate the thermal conductances of electrons and phonons across an Fe/MgO/Fe MTJ and reported an effective thermal conductivity of MgO of 0.15 W m$^{-1}$ K$^{-1}$ for a thickness of the MgO barrier of 1.15 nm. References [25] and [26] measured the TMS voltages of nanopillar and sputtered film MTJs, respectively, and determined the thermal conductivity of the oxide barriers by comparing with finite-element modeling. Reference [25] reported 0.005–0.2 W m$^{-1}$ K$^{-1}$ for the effective thermal conductivity of MgO, and Ref. [26] reported 5.8 W m$^{-1}$ K$^{-1}$ and 0.7 W m$^{-1}$ K$^{-1}$ as the upper limits for MgO and MgAl$_2$O$_4$ (MAO) tunnel barriers, respectively.

Despite the effort to determine the thermal conductance of thin oxide tunnel barriers, no direct measurement of temperatures in MTJs has been reported. In this work, we perform ultrafast thermometry on "half-MTJ" samples, which consist of only one FM electrode instead of two, in contact with an oxide tunnel barrier. The sample structure is Ru(50)/oxide(2)/FM/seed(5)/MgO where the number in parenthesis is the thickness of that layer in nm. For the oxide tunnel barrier, we study MgAl$_2$O$_4$ in addition to the more common MgO, as MAO is a promising candidate for tunnel barriers due to the similar spin-filter effect and smaller lattice mismatch with *bcc* magnetic metals, e.g., Fe, CoFe, and CoFeB, compared with MgO [27,28]. We chose a barrier thickness of 2 nm for both barrier materials, since in prior experiments we obtained the largest TMR values for this barrier thickness within a series of thickness-varied MTJs [29]. For the FM layer, we use Co or CoFeB, grown on top of a seed layer, Pt or Ta, respectively.

The optically thick Ru layer is heated by ultrafast laser pulses, and the subsequent temperature evolutions in the sample are observed by measuring time-domain thermoreflectance (TDTR) of Ru and time-resolved magneto-optic Kerr effect (TR-MOKE) of the FM layer. TDTR has been extensively used for studying heat transport in various materials at nanoscale [19,30]. However, the TDTR signal depends on the electron ($T_e$) and phonon ($T_{ph}$) temperatures as well as laser-induced strains within approximately the optical absorption depth. Thus, the interpretation of TDTR signals is straightforward only after the electrons and phonons reach thermal equilibrium near the surface. In this work, we propose to use an ultrathin magnetic layer as a thermometer as the TR-MOKE signal of the magnetic layer allows us to monitor the magnetic temperature of Co and CoFeB layers.

We independently investigate the magnetization dynamics of 6−10 nm thick Co and CoFeB single layers capped with a 2-nm-thick Pt layer, to characterize the response times of Co and CoFeB magnetizations to temperature changes. The relatively large thicknesses of 6−10 nm are needed to improve the sensitivities to the properties of Co and CoFeB but at the same time, they give rise to the in-plane directions as magnetic easy axis due to shape anisotropy. Thus, we use time-resolved quadratic magneto-optic Kerr effect (TR-QMOKE) [31,32] to observe the dynamics of the in-plane magnetization. From TR-QMOKE, we estimate the thermalization time of magnons with electrons, $\tau_{em} \approx C_m/g_{em}$, as 0.2 ps for Co and 2 ps for CoFeB, where $C_m$ and $g_{em}$ are magnon heat capacity and electron-magnon coupling parameters, respectively.

By combining TDTR and TR-MOKE on the half-MTJ samples, we are able to determine the value of $\Lambda_{oxide}$ of MgO and MgAl$_2$O$_4$ tunnel barriers. We note that $\Lambda_{oxide}$ is the effective thermal conductivity, which includes the thermal conductance of the Ru/oxide and oxide/FM interfaces in addition to the thermal conductivity of the thin oxide layer. We discuss the contributions of the interfaces of the tunnel barrier to $\Lambda_{oxide}$ and the non-equilibrium of electrons and phonons in metals near the tunnel barrier.

## II. EXPERIMENTAL METHODS

The samples for TR-QMOKE, Pt(2)/Co(10) and Pt(2)/Co$_{40}$Fe$_{40}$B$_{20}$(6.5), are deposited on *c*-cut sapphire substrates using a two-target DC magnetron sputtering deposition system at the University of Illinois. Throughout our discussion, the number in parenthesis is the thickness of the layer in nm. The Co(10) layer is deposited at 300°C to reduce the roughness and improve the crystallinity of the Co film, and capped with 2 nm Pt at elevated temperature, <300°C. Pt(2)/CoFeB(6.5)/sapphire is deposited at room temperature. The half-MTJ samples and control samples are deposited at room temperature using a multi-target magnetron sputtering system at Bielefeld University. The sample structures are Ru(50)/oxide(2)/Co(0.7)/Pt(5) and Ru(50)/oxide(2)/CoFeB(1)/Ta(5) on MgO(001) substrates. All samples are post-annealed at 360°C for 1 hour in an out-of-plane magnetic field of 0.7 T. We perform X-ray reflectivity and Rutherford backscattering spectrometry to confirm the layer thicknesses; vibrating sample magnetometer and alternating gradient magnetometer are used to identify the perpendicular magnetic anisotropy with the out-of-plane direction as the magnetic easy axis. See Fig. S1 for magnetic hysteresis loops of the half-MTJ samples.

Ultrafast thermal transport measurements are performed using a Ti:sapphire laser oscillator that generates a series of pulses at the repetition rate of 80 MHz with wavelength centered at 783 nm. The laser output is split into pump and probe beams having orthogonal polarizations and shifted wavelength spectra [33]. The pump beam is modulated by an electro-optic modulator at 11 MHz. The optical paths of the pump and probe beams are adjusted such that the beams are incident either on the same side or the opposite sides of the samples. The beams are focused by an objective lens and are incident normal to the sample surface. The pump and probe beams have the same $1/e^2$ radius of 5.5 µm for TDTR and TR-MOKE or 2.7 µm for TR-QMOKE. The cross-correlation of the pump and probe pulses is measured using a GaP photodiode via a two-photon absorption process and has a full-width-at-half-maximum of approximately 1.2 ps. The significant broadening in the laser pulses is caused by the electro-optic modulator and ultra-steep optical filters that we use to spectrally separate the pump and probe beams.

For TDTR, the intensity of the reflected probe beam is measured by a Si photodiode. For TR-QMOKE and TR-MOKE, the rotation of the polarization of the reflected probe is measured via balanced photodetection, i.e., a combination of a half-wave-plate, a Wollaston prism, and a balanced photodiode. To measure the Kerr ellipticity, the half-wave-plate in the MOKE detection setup is replaced with a quarter-wave-plate. The sample response synchronous to the modulation frequency of the pump is recorded by a lock-in amplifier. To improve the signal-to-noise ratio, data are averaged over 10-15 repetitions in TR-QMOKE and TR-MOKE measurements.

TR-QMOKE measurements [31,32] are performed with the orientation of the polarization of the pump and probe beams different from configuration used for TDTR and TR-MOKE. For TR-QMOKE on the Co and CoFeB samples having in-plane magnetic anisotropy, both the pump and probe are incident on the Pt(2) surface, and an in-plane magnetic field of ≈0.3 T is applied. We add a half-wave-plate before the objective lens to set the probe polarization at either +45° or –45° relative to the applied magnetic field. The difference of the two measurements with the probe at +45° and –45° gives the demagnetization signal detected by QMOKE; the sum of the two measurements corresponds to the out-of-plane component of the precessing magnetization detected via polar MOKE [31,32]. The half-wave-plate before the objective lens in the TR-QMOKE measurement also rotates the pump polarization to be at either –45° or +45° relative to the applied magnetic field and maintains the orthogonal polarizations of the pump and probe. The orthogonal polarization of the pump and probe suppresses an undesirable nonlinear optical effect, i.e., the optical Kerr effect, from contaminating the data during the temporal overlap of pump and probe [34].

The TR-MOKE measurement is performed at remanence on the samples with perpendicular magnetic anisotropy. To extract the TR-MOKE data, we take the difference between the Kerr rotation signals acquired at opposite magnetic polarities. The absolute values of static Kerr rotations ($\theta$) of Co and CoFeB samples are ≈0.5 mrad. We separately measure the temperature-dependence of complex Kerr rotation of the FM layers in the half-MTJ samples by using a photoelastic modulator and a heating stage in the range of $300 \leq T/\text{K} \leq 360$ (see Fig. S2). The temperature dependence of Kerr rotation, $|d\theta/dT|$, is $4\times10^{-6}$ K$^{-1}$ for Co(0.7) and $1.4\times10^{-6}$ K$^{-1}$ for CoFeB(1). The lower temperature coefficient of CoFeB might be due to the higher Curie temperature of CoFeB compared to Co, i.e., 750-1000 K for 1-nm-thick CoFeB. [35,36] and 600 K for sub-nm-thick Co [37].

### III. MAGNETIZATION DYNAMICS IN Co AND CoFeB MEASURED VIA TR-QMOKE

FM materials can provide a useful thermometer for studying laser-induced temperature evolution via TR-MOKE. This is possible when the laser fluence is small and the sample response to the laser excitation can be assumed to be linear. In this regime, the TR-MOKE signal can be approximated as linearly proportional to the magnon temperature of the FM [38]. For quantitative analysis of temperature evolutions, the knowledge of non-equilibrium energy transport properties of the magnetic materials is needed, such as electron-phonon ($g_{ep}$) and electron-magnon ($g_{em}$) coupling. Thus, we first perform TR-QMOKE measurement on the Pt(2)/Co(10)/sapphire and Pt(2)/CoFeB(6.5)/sapphire samples and separate demagnetization and precession behaviors, see Fig. 1 and Figs. S4-5. Then we compare the demagnetization data with a magnon temperature calculated by a three-temperature model (3TM) to determine two free parameters, $g_{ep}$ and $g_{em}$ of the FM layers. (See Supplementary Note S1 for the details of the 3TM; see Table S1 for materials parameters)

In Fig. 1, we compare the magnetization dynamics of Co and CoFeB with the best-fit of calculated temperature changes of electrons ($\Delta T_e$), phonons ($\Delta T_{ph}$), and magnons ($\Delta T_m$) at the center of the magnetic layers. Electrons are initially heated upon laser absorption and cooled by the exchange of thermal energy with magnons and phonons. The initial rise of $T_m$ is mostly determined by $g_{ep}$, $g_{em}$, $C_m$, and $\gamma_e$ of the FM layer (See Fig. S6 for sensitivities), in which $C_m$ and $\gamma_e$ are the magnon heat capacity and the electron heat capacity temperature coefficient, respectively. We use $C_m=0.02\times10^6$ J m$^{-3}$ K$^{-1}$ [39] and $\gamma_e=680$ J m$^{-3}$ K$^{-2}$ [40] from literature values for bulk Co and assume the same values for CoFeB. The two free parameters $g_{ep}$ and $g_{em}$ affect the magnitude and onset-time of the initial temperature-rise, respectively.

In Co, as shown in Fig. 1(a), electrons, phonons, and magnons are thermalized at about 3 ps. The plateau of the magnetic temperature after 3 ps implies that the deposited laser energy is confined in the Pt/Co metallic bilayer until the energy is transferred into the sapphire substrate at time delays $\geq$ 50 ps. (See Fig. S4(a)) The best-fit is obtained with $g_{ep}$(Co)=$2\times10^{18}$ W m$^{-3}$ K$^{-1}$ and $g_{em}$(Co)=$0.9\times10^{17}$ W m$^{-3}$. These values are similar to previous reports for FM metals, e.g.,

$g_{ep}=1\times10^{18}$ W m$^{-3}$ K$^{-1}$ and $g_{em}=1\times10^{17}$ W m$^{-3}$ K$^{-1}$ for Ni, and $g_{ep}=0.7\times10^{18}$ W m$^{-3}$ K$^{-1}$ and $g_{em}=0.6\times10^{17}$ W m$^{-3}$ K$^{-1}$ for FePt:Cu at 300 K [38].

The magnetization behavior of CoFeB as shown in Fig. 1(b), however, cannot be explained by a single set of $g_{ep}$ and $g_{em}$ values. The cooling of CoFeB is slower than the cooling of Co, such that the magnetic temperature reaches the plateau at about 10 ps. Modeling of the initial heating of $T_m$ yields $g_{ep}=1.1\times10^{18}$ W m$^{-3}$ K$^{-1}$ and $g_{em}=6\times10^{16}$ W m$^{-3}$, while modeling of the cooling of $T_m$ gives $g_{ep}=0.6\times10^{18}$ W m$^{-3}$ K$^{-1}$ and $g_{em}=1\times10^{16}$ W m$^{-3}$. Note that in this TR-QMOKE measurement, CoFeB stays in the linear response regime and the difference between the heating and cooling behaviors is not produced by changes in the magneto-optic constants induced by a laser pulse. This is supported by the fact that Kerr rotation and ellipticity show identical magnetization dynamics. (See Fig. S7) We do not yet understand the reason why CoFeB shows the different heating and cooling rates. We speculate that the non-thermal distribution of the optically excited electron-hole pairs may contribute to faster heating of the magnetic excitations of CoFeB in a manner that is not present in Co.

In the half-MTJ samples, the Co and CoFeB layers are 0.7 nm and 1 nm in thickness, respectively, much thinner than the other metallic layers. Thus, the overall temperature evolution is less sensitive to the thermophysical properties of the magnetic layers. The sensitivities of the magnetic temperature in Ru/MgO/Co/Pt are shown in Fig. S8. Phenomenologically, the contribution of the magnetic layer to the heat transport in the samples can be characterized by only two parameters: $g_{ep}$ and $\tau_{em}=C_m/g_{ep}$, where $\tau_{em}$ is the thermalization time of magnons due to the electron-magnon interaction. Moreover, in the half-MTJ samples, the magnetic layer is mainly heated by a phonon heat current through the oxide barrier rather than by direct laser excitation. Thus, the heating process is delayed relative to the laser excitation case and occurs at delay times of 10–3000 ps, as we discuss in more detail below. This time scale is much longer than the relaxation time scales of carrier non-equilibrium in Co and CoFeB. Thus, at time scales > 10 ps, the carrier coupling parameters of Co and CoFeB have a negligible effect on overall evolution of temperature in the samples. For CoFeB in the half-MTJ samples, we therefore use the parameters from the cooling behavior, i.e., $g_{ep}=0.6\times10^{18}$ W m$^{-3}$ K$^{-1}$ and $\tau_{em}\approx2$ ps. For Co, the parameters are consistent for ultrafast heating and cooling, i.e., $g_{ep}=2\times10^{18}$ W m$^{-3}$ K$^{-1}$ and $\tau_{em}\approx0.2$ ps.

Our approach includes the assumption that the non-equilibrium parameters of Co and CoFeB are similar for layers that are 10 nm and 1 nm thick. The Curie temperature of bulk Co and CoFeB is approximately 1300-1400 K; the Curie temperature is lower in thinner layers, approximately 600 K for sub-nm-thick Co [37] and 750-1000 K for 1-nm-thick CoFeB [35,36]. This suggests the magnon heat capacity ($C_m$) could be significantly higher in ultrathin layers than in bulk. On the other hand, prior work has suggested that the electron-phonon ($g_{ep}$) [41] and electron-magnon ($g_{em}$) [42,43] coupling parameters are increased for smaller thicknesses due to enhanced collision frequencies caused by boundary scattering. Thus, the changes in $C_m$ and $g_{em}$ for thinner magnetic layers are the opposite and mitigate the change in $\tau_{em}$. In our sample structures, a factor of five difference in $g_{ep}$ or $\tau_{em}$ of the Co(0.7) or CoFeB(1) layers changes the maximum magnon temperature by only 11%.

## IV.  TEMPERATURE EVOLUTIONS IN HALF-MTJS MEASURED VIA TDTR AND TR-MOKE

For the half-MTJ samples, we first perform TDTR measurements with both pump and probe incident on the surface of the optically thick Ru layer. The thermoreflectance of the Ru layer reports the surface temperature change of Ru following laser excitation. The ratio of the in-phase ($V_{in}$) and out-of-phase ($V_{out}$) TDTR signals at time delays < 50 ps is dominated by heat transport across the 50-nm-thick Ru layer; the ratio at time delays > 50 ps is dominated by heat transport from the Ru layer, through the oxide tunnel barrier, and into the MgO substrate, [30] and is shown in Fig. 2. Also shown are data for control samples without the oxide tunnel barrier. The TDTR ratio at time delays > 500 ps decreases less for the samples with the tunnel barriers. This implies the thermal conductance is smaller in the half-MTJ samples due to the additional layers between the Ru and substrate, i.e., the oxide tunnel barrier and its interfaces.

We model the samples as two layers, Ru and substrate, and use the analytic solution for TDTR signals in Ref. [30]. The free parameter is $G_{inter}$, the thermal conductance of all the intermediate layers between Ru and the substrate. The intermediate layers include the oxide tunnel barrier, FM layer, seed layer, and their interfaces. We model the interlayers as 1 nm in thickness but having the thickness-weighted heat capacities of all the layers. Since the metal layers, i.e., Co, CoFeB, Ta, and Pt, and the interfaces between the metal layers have high thermal conductivity and conductance, their contributions to $G_{inter}$ are negligible. $G_{inter}$ is mostly determined by the layers of the smallest thermal conductance. i.e., the effective thermal conductivity of the oxide tunnel barrier ($\Lambda_{oxide}$) and the thermal conductance of the interface between the seed layer (Pt or Ta) and the substrate ($G_{sub}$),

$$G_{inter}^{-1} \approx \left(\frac{\Lambda_{oxide}}{h}\right)^{-1} + G_{sub}^{-1} \qquad (2)$$

where $h$ is the thickness of the oxide tunnel barrier, i.e., $h=2$ nm. The best-fit of $G_{inter}$ ranges between 86–100 MW m$^{-2}$ K$^{-1}$, as shown in Table 1. The thermal conductance between Ru and substrate in the control samples can be approximated to $G_{sub}$, and the best-fit gives 190±30 MW m$^{-2}$ K$^{-1}$.

While the TDTR measurement gives only the sum of the reciprocal conductances, i.e., $\Lambda_{\text{oxide}}$ and $G_{\text{sub}}$, according to Eq. (2), the TR-MOKE measurement with the probe beam incident on the transparent MgO substrate probes the magnon temperature in the FM and allows us to separate the two parameters. We note that the quantitative analysis of TDTR is difficult to apply for the half-MTJ samples seen from the substrate side. This is because in the half-MTJ samples, several layers within the optical absorption length contribute to TDTR signal. These layers have different refractive indices as well as different electron and phonon temperatures, which complicate the interpretation of TDTR signal.

Figure 3 shows the TR-MOKE results of the samples with MgO tunnel barrier. (See Fig. S6 for MAO) The oxide barrier and the substrate are located on the opposite sides of the FM layer and affect the evolution of $\Delta T_m$ in the opposite ways: $\Delta T_m$ is higher for higher $\Lambda_{\text{oxide}}$ and lower $G_{\text{sub}}$. We define a sensitivity of $\Delta T_m$ to a material parameter $\alpha$

$$S(\alpha) = \frac{\partial(\Delta T_m)/\Delta T_{m,\text{max}}}{\partial \alpha / \alpha} \tag{3}$$

with $\Delta T_{m,\text{max}}$ as the maximum temperature rise of magnons. Figure 3(c) shows that the sensitivity to $\Lambda_{\text{oxide}}$ is positive and peaks at time delay $\approx$ 150 ps, while the sensitivity to $G_{\text{sub}}$ is negative and peaks at $\approx$ 300 ps. Figure 3(c) also shows that the sensitivities to the carrier coupling parameters, $g_{\text{ep}}$ and $\tau_{\text{em}}$ of CoFeB are negligible.

In the half-MTJ samples, most of the laser energy is absorbed by Ru as the Ru thickness is 50 nm and much longer than its optical absorption depth, $\approx$13 nm. (See Table S1) We calculate the optical absorption profiles by a transfer matrix method with complex refractive index of constituent materials, see Fig. S3 and Table S1. Only a small tail of the absorption profile lies in the FM and seed layers and causes ultrafast demagnetization in FM, as shown in Figs. S10-11. Taking Ru(50)/MgO(2)/Co(0.7)/Pt(4.4)/MgO in Fig. 3(a) as an example, the relative absorbance is 99%, 0.1%, and 1% for Ru, Co, and Pt, respectively. As we point out above, the electron heat current through the oxide tunnel barrier is negligible, $\approx 2$ MW m$^{-2}$ K$^{-1}$, and phonon heat transport dominates near the oxide barrier. For determination of $\Lambda_{oxide}$ and $G_{sub}$, we compare the TR-MOKE data of time delay between 10–3600 ps with 3TM calculations. We consider only the in-phase signal ($V_{in}$) of TR-MOKE as the out-of-phase signal ($V_{out}$) is small and taking a ratio of $-V_{in}/V_{out}$ significantly reduces the signal-to-noise ratio. To account for the divergence of the pump beam size, 16%, across the range of the linear delay stage, the in-phase voltage is multiplied by a factor of (1+0.16 $t_d$/3600), where $t_d$ is the pump-probe time delay in ps.

Figure 4 and Table 1 summarize our results for $\Lambda_{oxide}$ and $G_{sub}$ from TDTR and TR-MOKE measurements. The contour in Fig. 4(a) represents a set of values for $\Lambda_{oxide}$ and $G_{sub}$ for TR-MOKE data acquired on the Ru/MgO/Co/Pt sample that satisfies $\sigma^2 = 2\sigma_{min}^2$, where $\sigma^2$ is the sum of the squares of the residuals. The contour is limited by the two curves representing the range of $G_{inter}$=100±8 MW m$^{-1}$ K$^{-1}$ derived from the TDTR data. Figure 4(b) shows the range of $\Lambda_{oxide}$ and $G_{sub}$ of all four samples determined from both TDTR and TR-MOKE. We further restrict the range of $G_{sub}$ as 190±30 MW m$^{-1}$ K$^{-1}$ that is derived from the control samples having no tunnel barrier. $\Lambda_{oxide}$ and $G_{sub}$ of all four samples are consistent with one another within experimental uncertainty.

## V. DISCUSSION

The thermal conductance of interfaces ($G$) between different materials often plays a key role in heat transport on nanometer length scales [19]. At interfaces between a metal and a non-metal, heat transport is controlled by the phonon dispersion of the constituent materials and the transmission coefficient of phonons across the interface. Wilson *et al.* [20] showed that for clean and strongly-bonded interfaces between materials, the observed thermal conductance is approximately 40% of the maximum value of the conductance calculated for a transmission coefficient of unity, $G_{max}$, for the material that has the smaller value of $G_{max}$ between the two materials that make up the interface. This conclusion is similar to the prediction of the "metal irradiance" model recently described by Blank and Weber [44]. According to Ref. [20], a clean interface of Al/MgO has $G \approx 0.5$ GW m$^{-2}$ K$^{-1}$. The Debye temperatures of Co, Fe, and Al are similar, i.e., 460 K, 477 K, and 433 K, respectively [40], and the values of $G_{max}$ for these materials are also similar. Therefore, we expect that $G = 0.5$ GW m$^{-2}$ K$^{-1}$ provides a good estimate of the thermal conductance of interfaces between the metallic FMs (Co and CoFeB) and oxide tunnel barriers (MgO and MAO) in MTJs.

The effective thermal conductivity of an oxide tunnel barrier ($\Lambda_{oxide}$) should include the contribution from the two interfaces, i.e., Ru/oxide and oxide/FM, in addition to the thermal conductivity of the oxide. The thermal conductivity of MgO as a 2-nm-thick tunnel barrier would be further reduced compared to the thermal conductivity of a MgO thin film [16] of approximately 4 W m$^{-1}$ K$^{-1}$ due to boundary scattering. However, we expect the interfacial contribution is the limiting factor in the thermal transport. We estimate the upper limit of $\Lambda_{oxide}$ as $\approx 0.45$ W m$^{-1}$ K$^{-1}$ from our estimate of $G$(CoFeB/MgO)=0.5 GW m$^{-2}$ K$^{-1}$ discussed above and the thermal conductivity of the MgO thin film in Ref. [16]. In this limit, the thermal resistance ($h/\Lambda$) of the MgO layer contributes only 10% to the thermal resistance of the oxide tunnel barrier.

From the TDTR and TR-MOKE measurements on the half-MTJ samples, we obtain $\Lambda_{oxide}$ that are close to this upper limit of $\approx 0.45$ W m$^{-1}$ K$^{-1}$, as shown in Fig. 4 and Table 1. We note that $\Lambda_{oxide}$ of the half-MTJ samples in this study includes Ru/oxide and oxide/FM interfaces, instead of the two oxide/FM interfaces in typical MTJs. However, Ru has a Debye temperature of 555 K [40] comparable to the Debye temperatures of Co and Fe. Thus, we do not anticipate a significant difference between the effective thermal conductance of Ru/oxide/Co, Ru/oxide/CoFeB, and CoFeB/oxide/CoFeB.

Lastly, non-equilibrium between electrons and phonons exists near the oxide barrier, as shown in Fig. 3 and Fig. S9 and in accordance with Ref. [24]. We also present the temperature profiles in the half-MTJ structures at the delay time of 300 ps in Fig. 5, from $\Delta T_i(z,t)$ calculated by the 3TM (Supplementary Note S1). The length-scale of electron-phonon non-equilibrium in Ru can be estimated as $(\Lambda_e/g_{ep})^{1/2} \approx 8$ nm when $\Lambda_e \gg \Lambda_{ph}$ [45], which is shorter than its optical absorption depth, 13 nm. This implies the electrons and phonons are rapidly thermalized as they diffuse across the optically thick Ru layer. Figure 5 shows that the electrons and phonons are thermalized at the distance of 45 nm from the irradiated surface. It is also consistent with our TR-MOKE results on the half-MTJ samples at short delay times, i.e., the direct optical excitation of the FM layer is always more important, and the fast transport of photo-excited electrons in Ru is absent.

The electron-phonon non-equilibrium near the oxide barrier is caused by the imbalance between the electron and phonon currents through the oxide barrier and by the finite carrier coupling parameters of the metal layers in contact with the oxide barrier, Ru and FM. The thermal conductance of electrons through the oxide barrier is $G_e \approx 2$ MW m$^{-2}$ K$^{-1}$, as derived from tunneling electrical resistance, and is much smaller than the thermal conductance of phonons, $G_{ph} \approx 200$–300 MW m$^{-2}$ K$^{-1}$, as derived from $\Lambda_{oxide}$. This imbalance between electron and phonon transport at the metal-oxide interface creates electron-phonon non-equilibrium in the metal adjacent to the interface. The corresponding thermal conductance between electrons and phonons can be estimated as $G_{ep} \approx (g_{ep}\Lambda_p)^{1/2}$ if the metal layer is thicker than the non-equilibrium length-scale and $\Lambda_p \ll \Lambda_e$ [46]. If the metal layer is thinner, such as Co and CoFeB layers of $\leq 1$ nm thickness, we expect $G_{ep} \approx g_{ep}h$, where $h$ is the thickness of the FM layer, provides a good estimate, as we see below.

The energy exchange and transport near the oxide tunnel barrier can be described by a series of three thermal conductances: $G_{ep}$(Ru) near the Ru/MgO interface, $G_{ph}$(MgO) through the MgO barrier, and $G_{ep}$(FM) near the MgO/FM interface. The relative temperature differences as calculated by the 3TM in Fig. 5 are 7%, 81%, and 12% for ($T_e$–$T_p$) at Ru/MgO, phonons across the MgO barrier, and ($T_p$–$T_e$) at MgO/FM, respectively, in Fig. 5(a) with Co; 5%, 77%, and 20%, respectively, in Fig. 5(b) with CoFeB. Assuming $G_{ph}$(MgO)=250 MW m$^{-1}$ K$^{-1}$ from this work, we estimate $G_{ep}$(Ru)≈3-4 GW m$^{-2}$ K$^{-1}$, $G_{ep}$(Co)≈1.7 GW m$^{-2}$ K$^{-1}$, and $G_{ep}$(CoFeB)≈1.2 GW m$^{-2}$ K$^{-1}$. Thus, $G_{ep}$ near the tunnel barrier is only a factor of 5−12 larger than $G_{ph}$ of the oxide tunnel barrier. This results in the drop in the electronic temperature about 20−30% larger than the drop in the phonon temperature across the tunnel barrier.

The value of $G_{ep}$(Ru) derived from the phenomenological 3TM calculation agrees with the estimate of $(g_{ep}\Lambda_p)^{1/2} \approx 3$ GW m$^{-1}$ K$^{-1}$. For Co and CoFeB, the temperature evolutions are also affected by the Co/Pt or CoFeB/Ta interfaces. Note that we assume the phonon thermal conductance at the Co/Pt and CoFeB/Ta interfaces as 150 MW m$^{-1}$ K$^{-1}$ and this introduces a temperature drop of phonons at these interfaces, as can be seen in Fig. 5. For the ultrathin layers of Co and CoFeB, the estimates of $g_{ep}h \approx 1.6$ GW m$^{-1}$ K$^{-1}$ for Co and 0.6 GW m$^{-1}$ K$^{-1}$ for CoFeB are in good agreement with $G_{ep}$(Co) and $G_{ep}$(CoFeB), respectively, from the 3TM calculations.

## VI. CONCLUSIONS

We demonstrate ultrafast thermometry on half-MTJ samples, which consist of an oxide tunnel barrier sandwiched by an optically thick Ru layer and an ultrathin FM layer. We use the thermoreflectance of Ru and the MOKE of the FM layer as fast optical thermometers. We first characterize the non-equilibrium carrier coupling parameters of the FM thermometers, Co and CoFeB, using TR-QMOKE. The ultrafast heating and cooling rates of Co are described by consistent values of $g_{ep}$ and $g_{em}$, while heating and cooling of CoFeB require changes in the coupling parameters. We then determine the effective thermal conductivity of the oxide tunnel barriers of MgO and MAO by a combination of TDTR and TR-MOKE on the half-MTJ samples. The thermoreflectance of Ru allows us to determine the series thermal conductance of the oxide barrier and the interface between the metallic seed layer and the substrate; the MOKE of the FM layer allows us to separate the two contributions. We obtain the effective thermal conductivity of the oxide tunnel barriers as $0.4-0.6$ W m$^{-1}$ K$^{-1}$ and do not observe any systematic differences between MgO and MgAl$_2$O$_4$, or between Co and CoFeB samples. The effective thermal conductivity of the oxide layer is significantly lower than the thermal conductivity of thin sputtered films of MgO, $\approx 4$ W m$^{-1}$ K$^{-1}$, and is predominately limited by the two metal-oxide interfaces on either side of the oxide barrier. Moreover, electrons and phonons are not in thermal equilibrium near the oxide tunnel barrier, which must be considered for accurate assessment of spin phenomena in MTJs driven by a temperature gradient.

**ACKNOWLEDGEMENTS**

Pump-probe measurements of thermal transport and analysis by the three-temperature model were carried out in the Frederick Seitz Materials Research Laboratory Central Research

Facilities, University of Illinois, and were supported by MURI W911NF-14-1-0016. The development of ultrafast quadratic magneto-optic Kerr effect measurements was undertaken as part of the Illinois Materials Research Science and Engineering Center, supported by the National Science Foundation MRSEC program under NSF award number DMR-1720633. H. Jang acknowledges the fellowship from Kwanjeong Education Foundation of Korea. T. Huebner has been supported by the Deutsche Forschungsgemeinschaft (DFG) within the priority program Spin Caloric Transport (SPP 1538, KU 3271/1-1, RE 1052/24-2). We thank Günter Reiss for making available the laboratory equipment at Bielefeld University for sample fabrication and characterization and acknowledge Tristan Matalla-Wagner for help with the vibrating sample magnetometer.

**Figure 1.** Magnetization dynamics of (a) Co and (b) CoFeB measured by time-resolved quadratic magneto-optic Kerr effect (TR-QMOKE). Sample structures are (a) Pt(2)/Co(10)/sapphire and (b) Pt(2)/CoFeB(6.5)/sapphire. Black symbol is the data obtained by taking a difference of two TR-QMOKE measurements with probe polarizations at +45° and –45° relative to the external in-plane magnetic field, ≈0.3 T. Solid lines are the best fit of the temperatures of electrons (blue), magnons (red), and phonons (black) at the midpoint of the Co or CoFeB layer calculated by the three-temperature model. The absorbed fluences are (a) 1.2 J m$^{-2}$ and (b) 0.8 J m$^{-2}$.

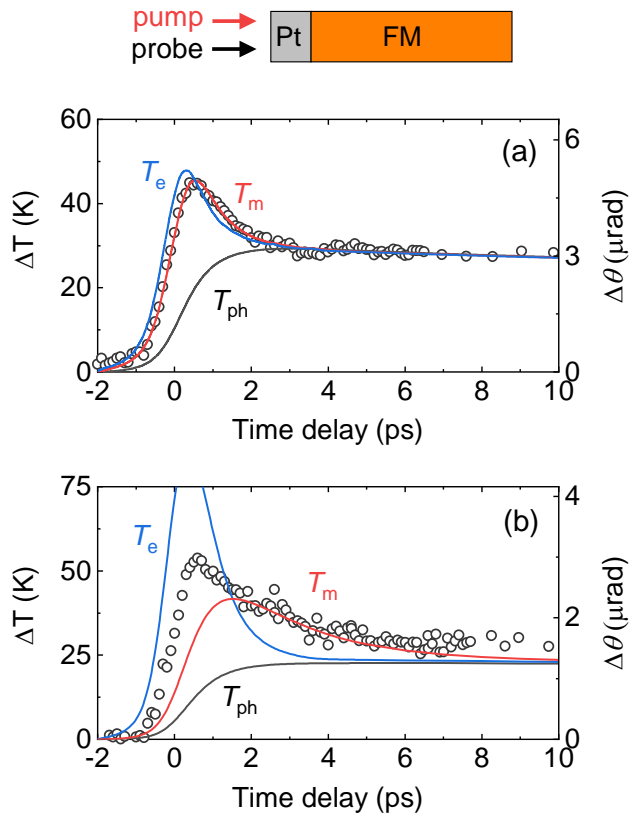

**Figure 2.** Time-domain thermoreflectance (TDTR) measurement with the pump and probe incident on the Ru surface of the half-MTJ samples with (a) MgO and (b) MgAl$_2$O$_4$ (MAO) tunnel barriers, and of control samples without tunnel barriers. Open symbols are measured TDTR data and solid lines are best-fit.

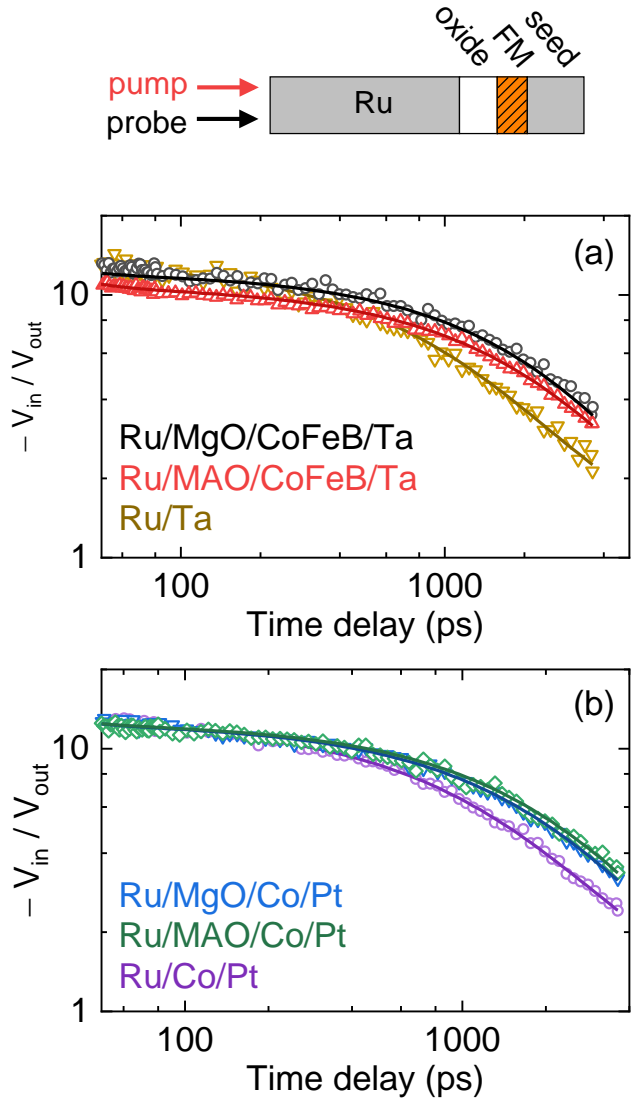

**Figure 3.** Temperature evolutions in (a) Co and (b) CoFeB of Ru/MgO/FM/seed/substrate samples when Ru is heated by the pump beam. Open symbols are TR-MOKE data measured with the probe incident on the MgO substrate. Solid lines are the best fit of the are the temperatures of electrons (blue), magnons (red), and phonons (black) in either Co or CoFeB calculated by the three-temperature model. (c) Sensitivity of magnon temperature in CoFeB to materials parameters for the sample configuration in (b). $\Lambda$, $h$, and $C_{ph}$ represent thermal conductivity, thickness and phonon heat capacity, respectively.

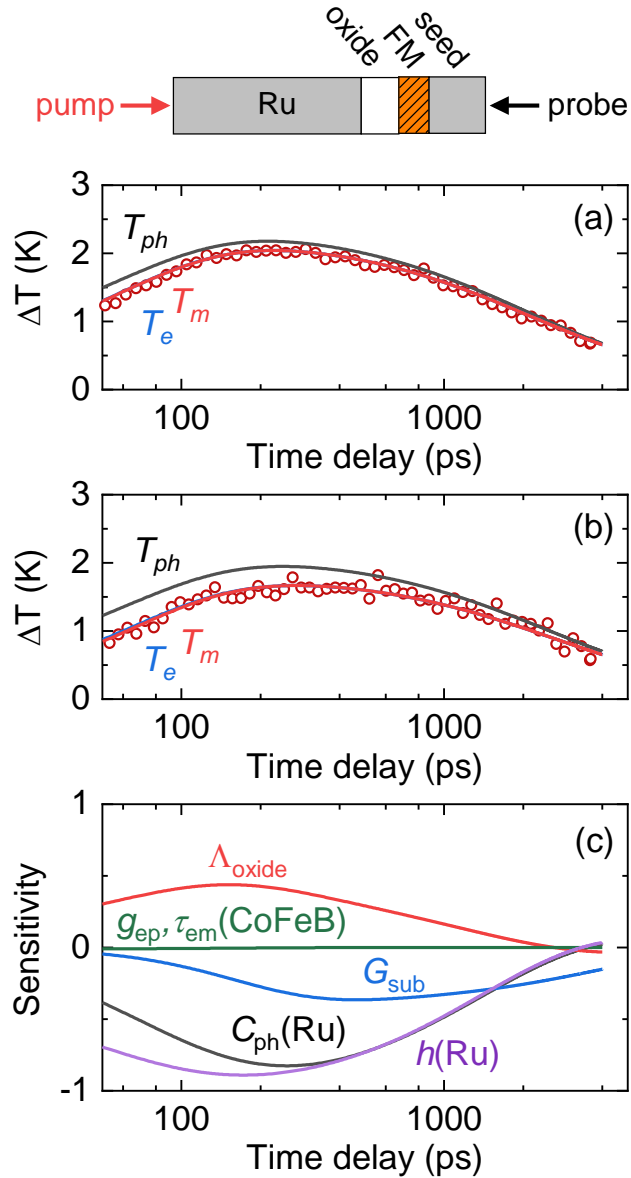

**Figure 4.** Best-fit of the effective thermal conductivity of oxide tunnel barriers ($\Lambda_{oxide}$) and the thermal conductance ($G_{sub}$) of the interface between the seed layer (Pt or Ta) and MgO substrate. (a) Contour (blue solid line) represents the range of $\Lambda_{oxide}$ and $G_{sub}$ determined from TR-MOKE on Ru/MgO/Co/Pt sample. Two curves (black dotted line) represents the range of $G_{int}$ determined from TDTR on the same sample. (b) Ranges of $\Lambda_{oxide}$ and $G_{sub}$ for all the samples derived from both TDTR and TR-MOKE. The shaded area represents the range of $G_{sub}$ determined by TDTR on the control samples without tunnel barriers.

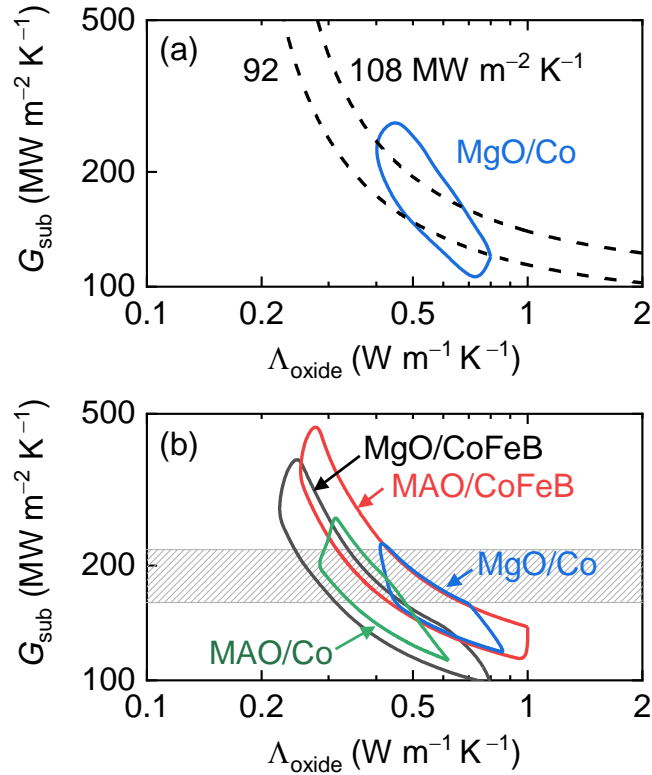

**Figure 5.** Temperature profiles at delay time of 300 ps in
(a) Ru(50)/MgO(2)/Co(0.7)/Pt(4.4)/MgO and (b) Ru(50)/MgO(2)/CoFeB(1)/Ta(4.6)/MgO with the pump beam incident on the Ru surface. The *x*-axis is the position with respect to the top surface of Ru. Solid lines are the temperatures of electrons (blue), magnons (red), and phonons (black) calculated by the three-temperature model. The temperatures of electrons and magnons appear overlapped (a) in Co and (b) in CoFeB as the differences are $< 0.1\%$.

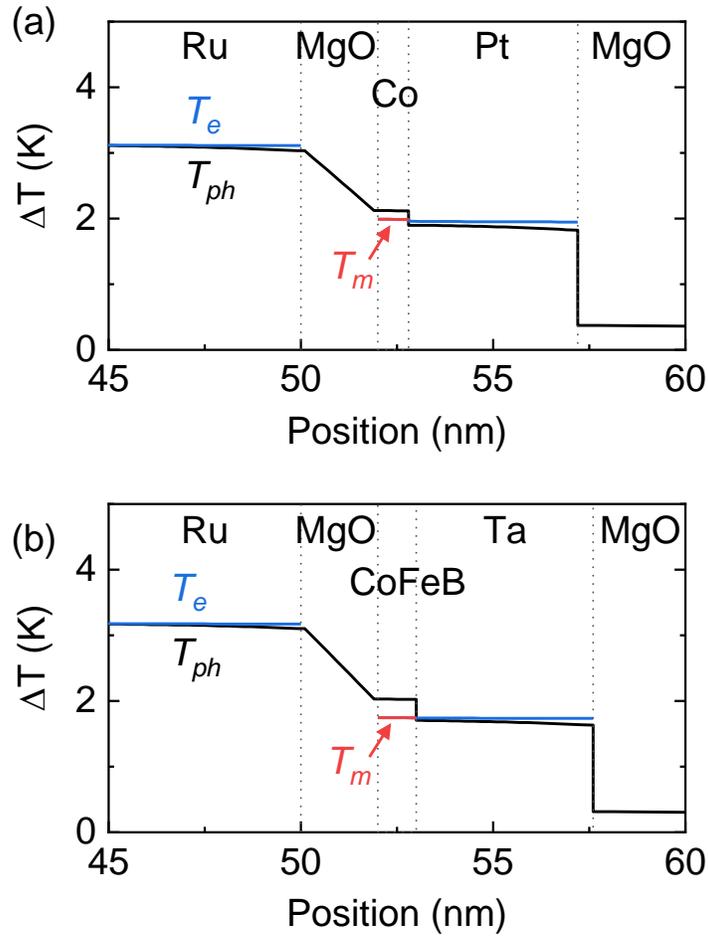

**Table 1.** Best-fit of thermal conductance of the intermediate layers between Ru and MgO substrate ($G_{\text{inter}}$) determined from TDTR and best-fit of effective thermal conductivity of oxide tunnel barriers ($\Lambda_{\text{oxide}}$) and thermal conductance of the interface between the seed layer and substrate ($G_{\text{sub}}$) determined from a combination of TDTR and TR-MOKE.

| Samples (oxide/FM) | $G_{\text{inter}}$ (MW m$^{-2}$ K$^{-1}$) | $\Lambda_{\text{oxide}}$ (W m$^{-1}$ K$^{-1}$) | $G_{\text{sub}}$ (MW m$^{-2}$ K$^{-1}$) |
|---|---|---|---|
| MgO/Co | 100±8 | 0.55±0.15 | 190±30 |
| MAO/Co | 90±8 | 0.4±0.1 | 190±30 |
| MgO/CoFeB | 86±8 | 0.38±0.13 | 190±30 |
| MAO/CoFeB | 98±8 | 0.5±0.2 | 190±30 |